\pdfoutput=1
\documentclass[aps,prb,twocolumn,letterpaper]{revtex4-1}

\usepackage{amsmath,amssymb}
\usepackage{txfonts}
\usepackage{textcomp}
\usepackage[pdftex]{graphicx,color}
\usepackage{braket}
\usepackage[pdftex,bookmarks=false]{hyperref}

\newcommand{\CeRu}{CeRu$_\text{2}$}

\begin{document}

\title{Effect of Geomagnetism on $^\text{101}$Ru Nuclear Quadrupole Resonance
Measurements of \CeRu}

\author{Masahiro Manago}
\email{manago@scphys.kyoto-u.ac.jp}
\author{Kenji Ishida}
\affiliation{Department of Physics, Graduate School of Science,
Kyoto University, Kyoto 606-8502, Japan}
\author{Tatsuma D. Matsuda}
\affiliation{Department of Physics, Tokyo Metropolitan University,
Hachioji, Tokyo 192-0397, Japan}
\author{Yoshichika \=Onuki}
\affiliation{Faculty of Science, University of the Ryukyus,
Nishihara, Okinawa 903-0213, Japan}

\begin{abstract}
	We performed $^{101}$Ru nuclear quadrupole resonance (NQR) measurements
	on the $s$-wave superconductor \CeRu{} and found oscillatory behavior
	in the spin-echo amplitude at the
	$\ket{\pm 1/2} \leftrightarrow \ket{\pm 3/2}$ transitions
	but not at the $\ket{\pm 3/2} \leftrightarrow \ket{\pm 5/2}$ transitions.
	The modulation disappears in the superconducting state
	or in a magnetic shield, which implies a geomagnetic field effect.
	Our results indicate that the NQR spin-echo decay curve
	at the $\ket{\pm 1/2} \leftrightarrow \ket{\pm 3/2}$ transitions
	is sensitive to a weak magnetic field.
\end{abstract}

\date{\today}
\maketitle

The spin-echo amplitude shows oscillatory behavior with respect to
the time interval between two pulses in nuclear magnetic resonance
(NMR) or nuclear quadrupole resonance (NQR) under some conditions.%
\cite{PhysRev.88.1070,PhysRev.97.1699,PhysRevLett.12.123,JPSJ.21.77,%
PhysRevB.53.14268}
There are various origins of the oscillation:
for example, an external or internal magnetic field in NQR
and indirect nuclear spin coupling cause such modulations.
These effects provide information on the electron system
surrounding the nuclear spin system.
In principle, a geomagnetic field can also be an origin of the modulation
of the spin-echo amplitude in NQR.
This effect seems to be negligibly small, but it actually causes
an appreciable modulation of the spin-echo decay curve as shown below.

We carried out $^{101}$Ru-NQR spin-echo decay measurements
on the conventional $s$-wave superconductor \CeRu\ ($T_\text{c} = 6.2$\,K)
to determine the $1/T_2$ behavior in the superconducting (SC) state.
$^{101}$Ru has nuclear spin $I = 5/2$, and two NQR peaks are observed
at 13.2 and 26.4\,MHz,
corresponding to the $\ket{\pm 1/2} \leftrightarrow \ket{\pm 3/2}$
and $\ket{\pm 3/2} \leftrightarrow \ket{\pm 5/2}$ transitions, respectively.
The $s$-wave SC character was verified from the exponential
temperature dependence of the nuclear spin-lattice relaxation rate
$1/T_1$ in the SC state,\cite{JPSJ.64.2750,Z.Naturforsch.51a.793}
although the Hebel-Slichter peak is considerably suppressed
owing to the anisotropy of the SC gap.\cite{JPSJ.67.2101}

A polycrystalline \CeRu\ sample was used for our measurements.
The SC transition at $T_\text{c} = 6.2$\,K was confirmed by
ac susceptibility measurement using an NMR coil.
All the measurements were performed in a glass dewar and the spin-echo
decay curve was obtained while varying the time interval $\tau$ between
the first ($\pi/2$) and second ($\pi$) pulses in an ordinary spin-echo method.
Any external magnetic field at the sample site, particularly
the geomagnetic field, was suppressed with a magnetic shield
made of a 1-mm-thick permalloy forming a half-open cylinder.

\begin{figure}
	\centering
	\includegraphics{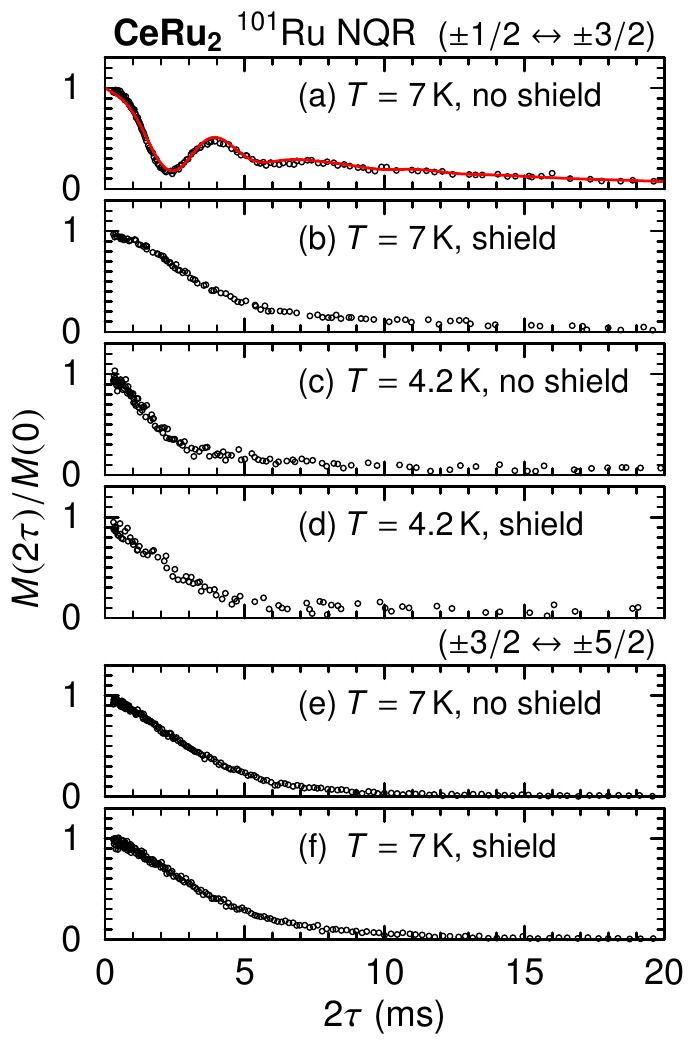}
	\caption{\label{fig:relaxation}(Color online)
	Spin-echo relaxation curves of $^{101}$Ru NQR on \CeRu.
	The curves were measured at the
	$\ket{\pm 1/2} \leftrightarrow \ket{\pm 3/2}$ line ($f = 13.205$\,MHz)
	at 7\,K (a) without and (b) with the magnetic shield.
	The curves of the same peak were also measured at 4.2\,K
	(c) without and (d) with the magnetic shield.
	The solid (red) line in (a) is the calculated curve obtained from
	Eq.~\eqref{eq:mexp} with $\mu_0H_0 = (3.93 \pm 0.11)\times 10^{-5}$\,T
	and the angle between $H_0$ and $H_1$ being 50\textdegree.
	The $\ket{\pm 3/2} \leftrightarrow \ket{\pm 5/2}$ curves
	($f = 26.41$\,MHz) at $T = 7$\,K (e) without and
	(f) with the magnetic shield do not show oscillatory behavior,
	which implies that the effect of the static magnetic field
	on this line is canceled by the spin-echo method.}
\end{figure}

Figure \ref{fig:relaxation} indicates the spin-echo decay curves
under various conditions.
Clear oscillatory behavior was observed in the spin-echo decay curve
measured at the NQR peak arising from the
$\ket{\pm 1/2} \leftrightarrow \ket{\pm 3/2}$ transitions above
$T_\text{c}$ (at 7\,K).
However, the oscillation disappears in the SC state (at 4.2\,K)
and in the measurement with the magnetic shield,
indicating that the oscillatory behavior is due to the geomagnetic field.
On the other hand, no oscillatory behavior was found at the NQR peak of
the $\ket{\pm 3/2} \leftrightarrow \ket{\pm 5/2}$ transitions even
in the measurement without the magnetic shield at $T = 7$\,K.

The above experimental results can be well understood
by quantum-mechanical analyses.
In the case of a half-odd ($I = 3/2,\ 5/2, \dots$) nuclear spin system,
the $\ket{I_z = \pm m}$ states are degenerate
in an electric field gradient (EFG) with axial symmetry.
The magnetic field lifts these degeneracies:
these states split into the $\ket{+m}$ and $\ket{-m}$ states for $m \ge 3/2$
and into $\ket{\phi_{+}}$ and $\ket{\phi_{-}}$ states, which are mixtures of
the $\ket{\pm 1/2}$ components, for $m = \pm 1/2$.
This removal of the degeneracy of the $m = \pm 1/2$ states is essential
for the oscillation of the spin-echo amplitude in the NQR peak arising from
the $\ket{\pm 1/2} \leftrightarrow \ket{\pm 3/2}$ transitions.
Note that the spin-echo method cannot cancel the effect of
a \emph{static} magnetic field in this peak.
Quantitatively, the spin-echo amplitude of the peak
of the $\ket{\pm 1/2} \leftrightarrow \ket{\pm 3/2}$ transitions
after time $2\tau$ from the first pulse in the static magnetic field $H_0$
for a single crystal sample is\cite{PhysRev.97.1699}
\begin{align}
	M(2\tau) &\propto \alpha \sin (2 \alpha \theta_{\text{FP}})
	\sin^2 (\alpha \theta_{\text{SP}}) \notag \\
	&\quad \times
	\Bigg[
		1 - \frac{2(f^2 - 1)}{f^2} \sin^2
		\left(
			\frac{f}{4} \Omega_0 \cos \theta_0 \cdot 2\tau
		\right) \notag\\
	&\quad\quad\quad\quad\quad \times \sin^2
		\left(
			\frac{3}{4} \Omega_0 \cos \theta_0 \cdot 2\tau
		\right)
	\Bigg],	\label{eq:m0} \\
\intertext{where}
	\alpha &\equiv \frac{1}{2}
	\left[
		I (I + 1) - \frac{3}{4}
	\right]^{1/2} \sin \theta_1, \notag \\
	f &\equiv
	\left[
		1 + \left(I + \frac{1}{2}\right)^2
		\tan^2 \theta_0
	\right]^{1/2}, \notag
\end{align}
$\tau$ is the time interval between two pulses, $\Omega_0 = \gamma H_0$ is
the Larmor frequency ($\gamma$ is the nuclear gyromagnetic ratio),
$\theta_0$ is the angle between the direction of the maximum principal
axis of EFG $V_{zz}$ and $H_0$, $\theta_1$ is the angle between $V_{zz}$
and the direction of the rf pulse field $H_1$,
$\theta_\text{FP} = \gamma H_1 t_\text{FP}$ ($t_\text{FP}$ is the time width
of the first pulse), and the same relation follows for $\theta_\text{SP}$.
Equation \eqref{eq:m0} is a generalization of the original formula
in Ref.~\onlinecite{PhysRev.97.1699} for an arbitrary half-odd spin.
Equation \eqref{eq:m0} holds only when the time constant
of free induction decay (FID), $T_2^*$, is sufficiently shorter than
$2\tau$ and the static magnetic field $H_0$ is homogeneous.
Other oscillatory components will appear
if $T_2^*$ is long, as shown in Ref.~\onlinecite{PhysRev.97.1699}.
In Eq.~\eqref{eq:m0}, the effects of any decay processes are neglected.
There is inhomogeneity of $H_0$ in a realistic system,
which causes damping of the oscillatory component.
Note that the oscillation vanishes when $\theta_0 = 0$ or $\pi$,
corresponding to the condition that $H_0$ does not mix the
$m = \pm 1/2$ states so that they form two independent systems.

On the other hand, it can be shown by a calculation similar to that
in Ref.~\onlinecite{PhysRev.97.1699} that there is no modulation of the spin-echo
relaxation curve in the other transition lines,
or $\ket{\pm m} \leftrightarrow \ket{\pm (m + 1)}$ for $m \ge 3/2$,
even if a static weak magnetic field is applied.
This is because two transitions, $\ket{m} \leftrightarrow \ket{m + 1}$
and $\ket{-m} \leftrightarrow \ket{-(m + 1)}$,
caused by the pulse field occur without tangling with each other,
since these two lines are independent as in the above case
of $\theta_0 = 0$ or $\pi$ in the $m = \pm 1/2$ states.
The absence of the oscillation observed in the
$\ket{\pm 3/2} \leftrightarrow \ket{\pm 5/2}$ transition lines
can be consistently understood with this scenario.

The experimental spin-echo intensity at $2\tau$ at
the $\ket{\pm 1/2} \leftrightarrow \ket{\pm 3/2}$
transitions, which shows the oscillatory decay behavior,
is reproduced with the following equation:
\begin{align}
	M_{\text{exp}}(2 \tau) &= M_0\, e^{-2 \tau/T_2}
	\Big[
		m_{\infty} + (m(2\tau) - m_{\infty})\, e^{-\delta \cdot 2\tau}
	\Big]. \label{eq:mexp}
\end{align}
$m(2\tau)$ is the integral of Eq.~\eqref{eq:m0} for the powder sample
[$m(0) = 1$], and the term $\exp(-\delta \cdot 2\tau)$
is introduced to take into account the damping of the oscillation.
The parameter $\delta$ corresponds to the inhomogeneity of the magnetic field.
Equation \eqref{eq:mexp} depends on the angle between $H_0$ and $H_1$,
$\theta$, in our case.
The calculated decay curve can be fit to the experimental data with
$\mu_0 H_0 \simeq 4\times 10^{-5}$\,T and $\theta = 50$\textdegree \
($\theta$ was fixed), and thus the oscillation
is ascribed to the geomagnetic field.

\begin{figure}
	\centering
	\includegraphics{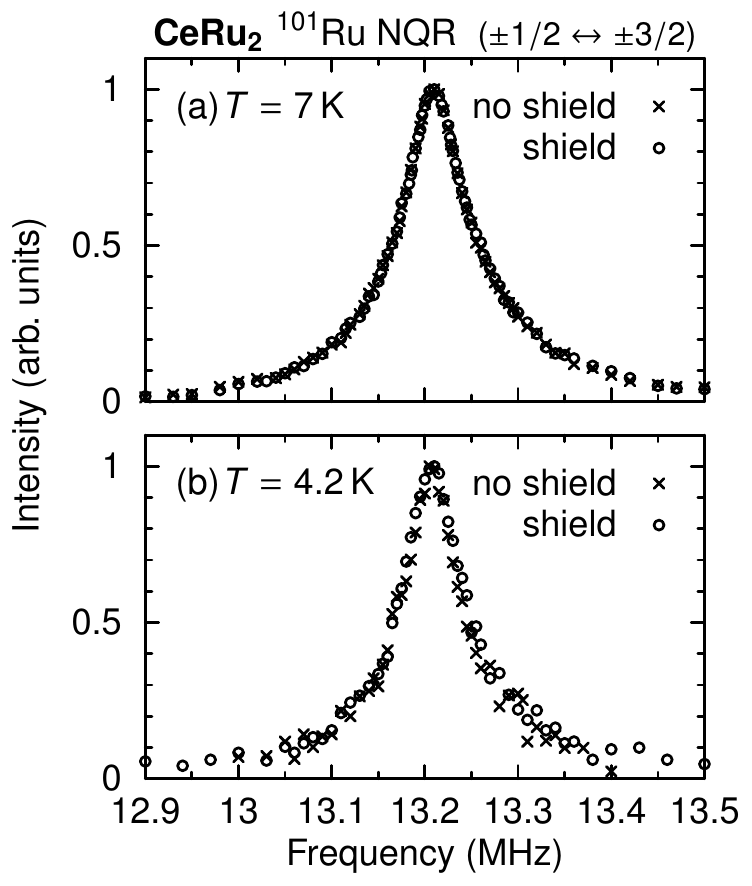}
	\caption{\label{fig:spectrum}
	$^{101}$Ru-NQR spectra of \CeRu.
	The spectra were measured at (a) $T = 7$\,K and
	(b) $T = 4.2$\,K with and without the magnetic shield.
	The FWHMs were $(97 \pm 2)$\,kHz both with and without the shield at 7\,K,
	and $(83 \pm 2)$\,kHz with the shield and $(88 \pm 4)$\,kHz
	without the shield at 4.2\,K.}
\end{figure}

In principle, this weak field should split or broaden the NQR line.
However, the additional broadening by the geomagnetic field is
estimated to be only $\simeq 4\times 10^{-4}$\,MHz, which is impossible
to detect experimentally, as shown in Fig.~\ref{fig:spectrum}.

Although we can eliminate the geomagnetic field effect, the value of $1/T_2$
was not well determined since the decay-curve behavior
strongly depends on the pulse condition.
This is because we could not optimize the pulse condition
owing to the relatively broad spectrum and small $\gamma$ of $^{101}$Ru.
A relatively narrow NQR spectrum, which can be excited by one $\pi/2$ pulse,
is necessary to obtain a reliable $1/T_2$ value.

In summary, we found that the spin-echo amplitude shows the oscillatory decay
at the NQR peak arising from the $\ket{\pm 1/2} \leftrightarrow \ket{\pm 3/2}$
transitions but not at the peak from
the $\ket{\pm 3/2} \leftrightarrow \ket{\pm 5/2}$ transitions on \CeRu{}
in the normal state.
Since the oscillatory behavior disappears in the SC state
and in the measurement with the magnetic shield,
this behavior originates from the geomagnetic field,
the magnitude of which is estimated from the frequency of the oscillation.
We show that the oscillatory behavior at the
$\ket{\pm 1/2} \leftrightarrow \ket{\pm 3/2}$ transitions
is useful for detecting such a small field
which does not cause an appreciable change in the NQR spectrum.

This work was partially supported by Kyoto Univ.\ LTM Center
and JSPS KAKENHI Grant No.\ 15H05745.

\bibliography{bibliography}
\end{document}